# Top-Down Behavioral Modeling Methodology Of A Piezoelectric Microgenerator For Integrated Power Harvesting Systems

Hela Boussetta (hela.boussetta@imag.fr), Skandar Basrour, Marcin Marzencki
Micro and Nano Systems Group - TIMA, 46 avenue Félix Viallet, 38000 Grenoble, FRANCE

*Abstract*-In this study, we developed a top/down methodology for behavioral and structural modeling of multi-domain microsystems. Then, we validated this methodology through a study case: a piezoelectric microgenerator. We also proved the effectiveness of VHDL-AMS language not only for modeling in behavioral and structural levels but also in writing physical models that can predict the experimental results. Finally, we validated these models by presenting and discussing simulations results.

I. INTRODUCTION AND MOTIVATION

The importance of wireless sensor networks is highlighted by the increasing number of applications, including disaster relief applications, intelligent building, facility management, environment control, machine monitoring and preventive maintenance.... Realizing such wireless sensor networks is a crucial step toward a deeply penetrating ambient intelligent concept. The power supply is a crucial system component since once the energy supply is exhausted; the node fails (Ref. [2]). To ensure truly long-lasting nodes, limited energy storage is unacceptable. Rather, Self Powered Micro Systems (SPMS) with their extended autonomy represent a promising answer. The challenge while designing such systems is to deal with the diversity of subsystems containing complex MEMS and different physical domains (mechanical and electrical). Since it is not possible to comprehend such complex systems in their entirety, we need to find methods of dealing with this complexity. In this work, we adopted a top/down methodology for behavioral and structural modeling of multi-domain microsystems. Then, we verified the efficiency of such a methodology by modeling and simulating an integrated power harvesting circuit.

II. MODELING MULTI-DOMAINS SYSTEMS

Several definitions of a model can be found in the literature. One possible definition is the following: "A model represents that information which is relevant and abstracts away irrelevant detail" (Ref. [3]). A direct consequence of this definition is that a system can be represented by several models depending on aspects the designer focus on. Usually, three aspects of modeling are considered: function, structure and geometry. The most abstract model considered in this work is the functional level where the system is described by is function but information about the implementation of this function is not considered. The structural model describes a system as a composition of subsystems connected together. Details about the geometrical and physical parameters of the system are only considered in the physical layer.

The chosen modeling language in this work is VHDL-AMS. Actually, this language offers facilities for describing structure and function of systems in physical domains.

III. PIEZOELECTRIC MICROGENERATOR MODELING

The purpose of this section is to establish a generic library of piezoelectric microgenerators in different abstraction levels. Depending on the wanted compromise time/precision, the user can make its choice.

*A. A functional transducer model*

The simplest and probably the most abstract model of a transducer is an equivalent electrical circuit composed of an ideal sinusoidal current source, in parallel with a capacitor Cp. This model represents electrical characteristics of the microgenerator. The current amplitude ip depends of the amount of mechanical vibrations but still relatively insensitive to external charge. However, it is not suitable to our study case since the coupling effect is considerable.

*B. A structural piezoelectric microgenerator model*

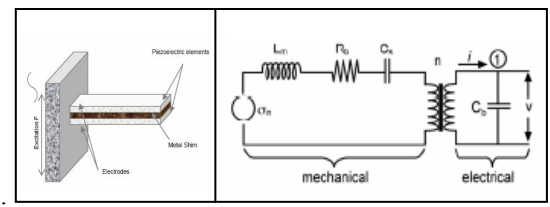

Fig. 1. Structural bimorph piezoelectric microgenerator model

The structural model described in Fig. 1. is developed by Ref. [4]. As shown in the left side of Fig. 1., the structure is a two layer bender mounted as a cantilever beam. The equivalent inductance represents the generator inertia. The mechanical stiffness is represented by the equivalent capacitance CK and σin represents the stress resulting of input vibrations. The transformation ratio n represents the coupling effect, Cb is the piezoelectricity capacitance and v(t) is the output voltage. The VHDL-AMS model is simple, doesn't require a lot of time to be written since it's obtained

 

by a simple connection of basic components (current source, inductor, resistance, capacitor and transformer). However, physical aspects of the system are not considered.

The same model can be described by analytical equations obtained from constitutive piezoelectric and circuit equations. The comparison between these two models is done in section IV.

### C. Physical piezoelectric microgenerator model

- **Piezoelectric microgenerator 1D model**

The target of this simple model is to validate our approach for transduction modeling. The whole system is subjected to mechanical vibrations y(t).

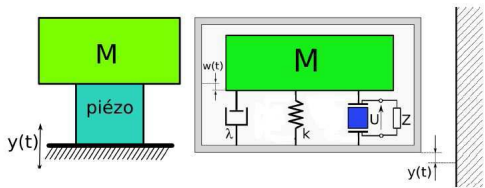

Fig. 2. Piezoelectric transducer model

The system is represented in Fig. 2. where:
– w(t) is the displacement of the seismic mass M.
– The stiffness K is calculated from the material stiffness $c_{33}$ and device dimensions.

Considering that this coupled electromechanical structure can be modeled as a damped harmonic oscillator, the analytical model is directly deduced from the Newton's second law of motion and the constitutive piezoelectricity equations.

The deduced differential equations are then correctly incorporated into a VHDL-AMS architecture.

This model was kept intentionally simple to focus on the functionality of the piezoelectric transduction. It is simple enough to have fast simulations as will be proven in section 5. This model is perfect for a first validation but it is neither accurate nor predictive.

- **Piezoelectric microgenerator 3D model**

The micropower generator is based on the structure illustrated in Fig. 3. It is based on a cantilever beam of length Lp, width Bp and the thickness HP on which a thin piezoelectric layer is deposited. A big seismic mass of length LM, width BM and thickness HM is attached at the end of the beam. This seismic mass is used to decrease the resonance frequency and increase the harvested energy. An applied acceleration induces the displacement of the mass and the deformation of the beam. Therefore, the beam apply constraints on the piezoelectric layer whose generate electric charges. Thanks to the proposed dimensions of this mass, we can tune the resonant frequency of the generator.

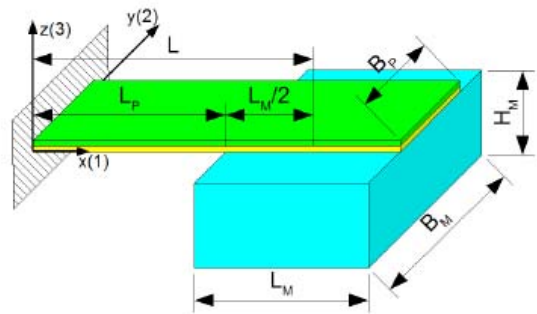

Fig. 3. Schematic of the structure modeled for the physical model

Several important considerations are taken into account especially the important size of the mass compared to the one of the beam, the rigidity of the mass and its rotational inertia. It is also important to consider that the acceleration is applied to the mass and not to the end of the beam. The impact of the electrical and mechanical properties of materials was introduced in the expression of the effective parameters deduced from boundary conditions. Additional information about the description of the model can be found in reference Ref. [1].

### IV. SIMULATIONS RESULTS

### A. Comparison between analytical and structural model results

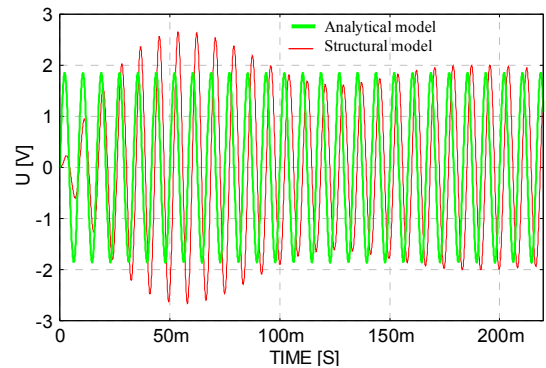

Fig. 4. comparison between structural and analytical model of bimorph piezoelectric model

Fig. 4. shows a comparison between analytical and structural models. The two curves are similar. The difference between the two models become from the abstraction done in the analytical model. In fact, the analytical model takes into account only the steady state while structural model consider both transient and steady state.

### B. Impact of physical properties study

First, we have studied the impact of the piezoelectric properties of the material used on the model by keeping the same dimensions of both devices and just changing the piezoelectric material layer. The used piezoelectric materials can be either Aluminium Nitrite (AlN) or Lead Zirconium Titanate (PZT) thin layers.

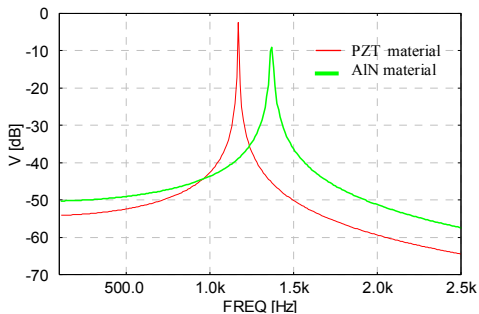

Fig. 5. Comparison between the piezoelectric microgenerator model based on PZT material and the one based on AlN material keeping the same mass dimensions

To do so, we just had to substitute the generic parameters responsible of physical properties of the used material (PZT) in the entity declaration of the previous of the VHDL-AMS model by AlN ones. The used stimulus is 1g acceleration.

As shown in For the AlN material, we have noted lower amplitude and higher resonant frequency. This result was expected because of the poor coupling coefficient of AlN material compared to PZT one.

However, microfabrication process in case of AlN by sputtering techniques is easier than for PZT ones. The deposition of AlN is relatively simple, compatible with CMOS process and does not require post process polarization. For that reasons, we decided to investigate structures based on this material.

We have changed the dimensions of the structure and analyzed the output voltage produced. We used twice the dimensions for the mass (800μm by 800μm) with a SOI wafer (525 μm thick). For acceleration stimuli of 1g amplitude, we obtained the output voltage versus time reported in Fig. 6. We can successfully harvest 1.8V across the electrodes in open circuit.

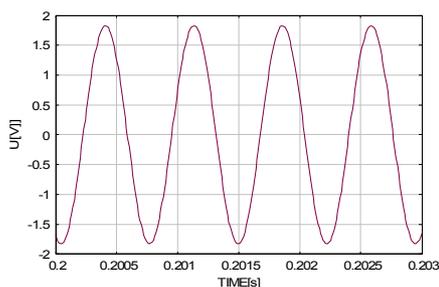

Fig. 6. Output voltage at resonance at micropower generator (with ALN as piezoelectric layer and modified dimensions of the mass) versus time

## V. DISCUSSION

The various models of the piezoelectric microgenerator developed in this paper demonstrate clearly that the choice of the degree of abstraction is closely related to the wanted performance: behavior virtual IP for a functional verification early in the design process which is very useful for simulation of complex and big designs or a predictive model comparable to experimental results.

Since all models are generic interfaces lists, several analysis possibilities can be performed just by changing geometric and physical parameters. This kind of tests is very important in the design process since it offers to the designer the opportunity to optimize his model early in the design process. In section C, we studied the impact of geometric properties of the device. We demonstrate by changing the dimensions of the structure and the piezoelectric material, we can harvest a great voltage with easier micro fabrication process. The user still has the possibility to experiment other materials and other geometric dimensions in order to optimize his design without dealing with complicated details of the model. Finally, we had performed global simulations using the MEMS model connected to an electrical circuit based on ultra low threshold voltage diodes used to boost and rectify the weak amplitudes AC signals delivered by such generators (voltages often inferior to 200 mV) (Ref. [5]).

## VI. CONCLUSION AND FUTURE WORK

The reusable aspect of our models offers to designers the possibility to select and use their suitable configurations without having to understand the details of blocks just by changing some parameters. Indeed, the piezoelectric generator model remains valid for other materials and the voltage multiplier circuit is extensible to other technologies. We also proved the effectiveness of VHDL-AMS for modeling behavioral, structural and physical level that can predict the experimental results. Thus, because the models respect energy conservation laws, we had performed global simulations using the MEMS model connected to an electrical circuit used to manage power delivered by the microgenerator. Further work must be done to take into account the damping effect not considered in this work; only viscous damping was considered.

ACKNOWLEDGMENT

The " Region Rhônes Alpes" and the University Agency for Francophony (AUF) are gratefully acknowledged for their financial support.

BIBLIOGRAPHY

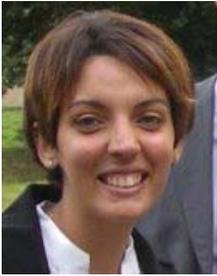

Hela BOUSSETTA was born in 1980 in Tunis, Tunisia. She is an electronical engineer and received a master degree in NTSID (New Technologies Of Computer Systems Dedicated Technologies) in 2004 from the ENIS (Tunisia). The master internship was done at STMicroelectronics-Tunis. During her internship, Ms Boussetta participates in developing STBus Transaction Level Models (TLM) using SystemC2.0. Currently, Ms Boussetta is being PhD student at TIMA Laboratory-Grenoble, where her researches focus on multi-domain modeling using VHDL-AMS and SPICE.